\documentstyle[osa,manuscript]{revtex}  
\begin{document}                
\title{Interpretation of the topological terms in gauge system}
\author{Zhong Tang * ~ and ~ David  Finkelstein **}

\address{School of Physics, Georgia Institute of Technology, 
Atlanta, Georgia 30332-0430}

\maketitle

\begin{abstract} 

We provide an alternative interpretation for the topological terms
in physics by investigating the low-energy gauge interacting system.
The asymptotic behavior of the gauge field at infinity
indicates that it traces out a closed loop 
in the infinite time interval: $(-\infty,~+\infty)$. 
Adopting Berry's argument of geometric 
phase, we show that the adiabatic evolution
of the gauge system around the loop results in an additional
term to the effective action:
the Chern-Simons term for three-dimensional spacetime, 
and the Pontrjagin term for the four-dimensional spacetime. \\

\noindent PACS number(s):  03.65.Bz, 11.15.-q, 02.40.+m.\\

\noindent 
Electronic addresses: 

 *gt8822b@prism.gatech.edu, 
 **david.finkelstein@physics.gatech.edu.

\end{abstract}

\newpage

As the property uniquely appearing in quantum theory,
topology has been acquiring much attention
in both condensed matter and particle physics.
The topological term for example the Pontrjagin term
was first introduced in gauge field theory
to resolve the anomaly in triangle graphs
that breaks down the usual Ward-Takahashi identity of the  
chiral current \cite{anomaly}.
It was found later that the topological term 
yields the $instanton$ solution
\cite{instanton}, which gives the local minima of the 
Yang-Mills gauge field action
and  brings in an interesting connection 
between particle physics and spacetime topology.
Recently, Witten \cite{witten} introduced topological 
quantum field theory, which emerges as possible realization
of general coordinate invariant symmetries.

One notices that in the above quantum theories, the topological 
term is always put in the action $by~ hand$, not based on a 
dynamical consideration. In this Letter,
we attempt to give an interpretation of the topological term 
alternatively by exploring the implication of geometric phase 
in the path integral formalism of low-energy gauge interaction. 

As a powerful method of quantizing quantum theory,
the path integral method was developed by Feynman 
\cite{feynman}, 
based on Dirac's intuition \cite{dirac}  that the transition 
amplitude of a quantum system between two states $|\alpha, 
t \rangle$ and $|\alpha', t' \rangle$ is proportional 
to the phase factor in terms of 
the classical action of the same system ($\hbar=c=1$):
\begin{equation}
\langle\alpha', t'|\alpha, t \rangle\propto e^{iS(t', t)}.
\end{equation}
The path integral has developed into 
a functional integral approach to quantum field theory,
which not only yields a 
simple, covariant quantization of complicated
systems with constraints, such as gauge theory,
but also leads to a deep understanding to some basic 
assumptions of quantum theory. 
 
However, the ordinary path integral method
needs improving in dealing with some quantum 
mechanical systems, for  example, the adiabatically evolving 
systems containing geometric phases. To see this clearly, 
we recapitulate the basic idea of the geometric phase
proposed by Berry \cite{berry}. For a system in which the 
Hamiltonian 
$H$ evolves adiabatically with parameters ${\bf R}\equiv{\bf R}(t)$,
and has a  $discrete$ spectrum: 
$H({\bf R})|n,{\bf R} \rangle = E_{n}({\bf R}) |n,{\bf R}\rangle$.
The state  of the system, determined by
the Schr$\ddot{o}$dinger equation:
\begin{equation}
H({\bf R})|\Psi_{n}(t) \rangle=i\frac{\partial}{\partial t}
|\Psi_{n}(t) \rangle,
\end{equation}
is solved as: $|\Psi_{n}(t) \rangle=e^{i\left[\gamma_{n}(t)+
\gamma'_{n}(t)\right]} |n,{\bf R}(t)\rangle$,
where $\gamma_{n}(t)=i\int_{0}^{t}\langle n,{\bf R}|\partial_{t'} 
|n,{\bf R}\rangle d t' $, and  
$\gamma'_{n}(t)=-\int_{0}^{t}
E_{n}({\bf R}(t')) d t'$.
If ${\bf R}$ executes a closed loop:  ${\bf R}(T)={\bf R}(0)$, 
$\gamma_{n}(T)$ is expressed alternatively as:
\begin{equation}
\gamma_{n}(T)=\oint {\bf A}({\bf R})\cdot d {\bf R}, 
\end{equation}
where ${\bf A}({\bf R})\equiv i\langle n,{\bf R}|{\bf \nabla}_{{\bf 
R}}
|n,{\bf R}\rangle$ is called Berry's potential. 
It was shown by Simon \cite{simon} that the above 
$\gamma_{n}(T)$ is attributed to the holonomy 
in the parameter space,  thereby called $geometric ~phase$.

For the above cyclic evolutionary system,
the transition amplitude of the states after
the parameter  ${\bf R}$  traces out a closed loop is
obtained to be:
\begin{equation}
\langle \Psi_{n}(T)|\Psi_{n}(0)\rangle=
e^{i\left[\gamma_{n}(T)+\gamma'_{n}(T)\right]}.
\end{equation}
Comparing this result with Eq. (1), we see immediately that 
the classical action includes the dynamic phase only.
This can be understood from two aspects: 
(i)  The topological structure of quantum theory does 
not show up in the classical dynamics;
(ii) Geometric phase as the quantity of one-order 
time derivative does not contribute to the usual 
Lagrangian equation of motion and the classical action either.

Let us further look at the gauge system with infinite number 
of degrees of freedom, which is usually treated by the perturbation 
theory in the interaction picture based on the adiabatic 
approximation.  
To be consistent with the boundary requirement of spacetime 
topology, the gauge field ${\bf a}({\bf x},t)$ (the temporal
gauge is chosen: $a_{0}({\bf x},t)=0$) 
generally has the following asymptotic 
behavior at infinity \cite{nash}: 
\begin{equation}\left\{
\begin{array}{l}
{\bf a}({\bf x},t)\rightarrow 0, ~~~\mbox{for}~~~
 t\rightarrow\pm \infty;\\[.12in]
{\bf a}({\bf x},t)\rightarrow {\bf \nabla}g(x), ~~~\mbox{for}~~~
|{\bf x}|\rightarrow \infty.
\end{array}\right.
\end{equation}
These conditions further suggest
that for sufficiently low-energy gauge interaction  such that
the creation and annihilation of particles are negligible, 
the gauge field  ${\bf a}({\bf x},t)$
can be taken as the parameter space. Then
the gauge system undergoes a $cyclic~ evolution$ from 
$t\rightarrow -\infty$ to $t\rightarrow +\infty$.
A question rises immediately:

How to take into account the effect of cyclic evolution 
in the above low-energy gauge interacting system?

Before a further discussion to the above question, 
we would like to make a digression
to mention a recent discovery by Newton \cite{newton}:
For a quantum mechanical system with continuous spectra 
for instance the scattering case, 
Newton introduced a so-called noninteraction picture to
show that the system presents a cyclic change with 
time $t$  from $-\infty $ to $+\infty$.
The geometric phase factor is then proved to be nothing 
but the $S$ matrix. We know that the $S$ matrix in the scattering 
theory is easily formulated in terms of the path integral.
This implies that the geometric phase factor can be described 
by  the path integral. We infer further that for the 
above gauge system presenting a cyclic evolution,
Berry's argument on geometric phase probably shows its
own $effect$ in the path integral formalism of the system.

The purpose of this Letter is to approach the above $effect$. 
We first explain why we prefer the gauge system:
(i) The importance of gauge theory in the 
description of the elementary particles and forces;
(ii) It has been shown that gauge structure
appears in the geometric phase \cite{wilczek}. 
Therefore, it would be interesting to investigate the role  
this gauge structure plays in the gauge  theory.
Without loss of generality, we consider the Abelian
gauge interacting system with the action:
\begin{equation}
S=\int d^{4}x\left[i\bar{\Psi}(\gamma^{\mu}
D_{\mu}+im)\Psi -{1\over{4}}F_{\mu\nu}F^{\mu\nu}\right],
\end{equation}
where $\Psi(x)$ is the fermion field with mass $m$, 
$D_{\mu}=\partial_{\mu}+iea_{\mu}$ and $F_{\mu\nu}=
\partial_{\mu} a_{\nu}-\partial_{\nu}a_{\mu}$. 
We can not directly follow either
Berry's calculation or  Newton's noninteraction
picture,  because: 
(i) the above gauge system is unbounded and has continuous 
spectrum, (ii) quantum fields are described  by infinite 
dimensional Hilbert space.
In the present Letter, we are not going to get involved in
the complication of the path integral formalism of the 
adiabatically evolving system. Instead, we provide
a simple way outlined as follows.

As we know that the gauge field $a_{\mu}(x)$ in above action
is in fact a quantum object. However,
if we take into account the geometric phase of the gauge system, 
$a_{\mu}(x)$ should be taken as the classical parameter 
(extension of geometric phase from classical to quantum
will be discussed later). 
For this purpose, we consider only the case of sufficiently 
low-energy interaction, where the creation and annihilation of 
particles are negligible, and the gauge field can be treated as 
the parameters for the first order approximation. 
Under the adiabatic approximation, 
the equation of motion of fermion field is in nearly free-form, 
which is conveniently handled in momentum space. Moreover,
since the gauge field is taken as the parameter, 
each mode of the fermion field in momentum space is  
independent of others,
thus can be effectively treated by quantum mechanics. 
For simplicity in description, 
we neglect the mass term of the fermion field. 
Then the Hamiltonian of each mode in momentum space 
is written as:
\begin{equation}
h({\bf a})=-\alpha_{i}(p_{i}+e a_{i}),
\end{equation}
where $\alpha_{i}=\sigma_{i}\bigotimes\sigma_{1},
i=1,2,3$,  $\sigma_{i}$ are the Pauli matrices,
and the gauge condition $a_{0}=0$ is chosen.
With these preparations, we now compute the total effect
of the geometric phase in the above gauge system 
through two steps:

First, since  the gauge field in Eq. (7) satisfies the asymptotic 
condition: ${\bf a}(t\rightarrow -\infty)=
{\bf a} (t\rightarrow +\infty)=0$,
namely, the gauge field executes a closed loop with 
respect to the infinite time interval: $t\in( -\infty,~ +\infty)$, 
each mode of the fermion field in 
momentum space will acquire a geometric phase after the gauge 
field traces out the loop. Fortunately, 
this geometric phase can be directly obtained by 
applying Berry's method to $h({\bf a})$. 
For the eigenvalue equation in parameter space:
$h({\bf a})|\psi({\bf a})\rangle=
E({\bf a})|\psi({\bf a})\rangle$,
we choose only the positive energy 
solution: $E_{+}({\bf a})=\sqrt{({\bf p}+e{\bf a})^{2}}$,
since the states are unbounded,
where eigenkets are doubly-degenerate:
\begin{equation}
|\psi_{+}({\bf a})\rangle=\frac{1}{\sqrt{2}}\left(
\begin{array}{c}
\sin\theta e^{-i\phi}\\
-\cos\theta\\
0\\
-1
\end{array}\right),
~~~|\psi_{+}({\bf a})\rangle'=\frac{1}{\sqrt{2}}\left(
\begin{array}{c}
-\cos\theta\\
-\sin\theta e^{i\phi}\\
1\\
0
\end{array}\right),
\end{equation}
where $\theta $ and $\phi$ are polar and azimuthal angles 
of the vector ${\bf q}\equiv {\bf p}+e{\bf a}$, respectively.
Then for each mode in momentum space,
the geometric phase of the ket 
$|\psi_{+}({\bf a})\rangle$ is \cite{nonabel}:
\begin{equation}
\gamma_{+}({\bf p})=e\int\limits_{-\infty}^{+\infty} dt
~A_{i}({\bf p}+e{\bf a}) ~\dot{a}_{i},
\end{equation}
where $
A_{i}({\bf p}+e{\bf a})=i\langle\psi_{+}({\bf a})|
\frac{\partial}{\partial q_{i}}|\psi_{+}({\bf a})\rangle
\equiv A_{i}$ are given by:
\begin{equation}
A_{1}=-\frac{q_{2}}{2{\bf q}^{2}},
~~~A_{2}=\frac{q_{1}}{2{\bf q}^{2}}, ~~~A_{3}=0,
\end{equation}
and satisfy the relation
\begin{equation}
\frac{\partial A_{j}}{\partial q_{i}}-
\frac{\partial A_{i}}{\partial q_{j}}=
\frac{q_{3}}{{\bf q}^{4}}
q_{k} \epsilon_{ijk}+\pi\delta(q_{1}, 
q_{2})\Delta(q_{3})\epsilon_{ij3},
\end{equation}
where the function $\Delta(q_{3})=1$ for $q_{3}=0$;
otherwise,  $\Delta(q_{3})=0$.  
Under the low-energy approximation, we expand 
$A_{i}({\bf p}+e{\bf a})$ with
respect to the gauge field up to the first order:
$A_{i}({\bf p}+e{\bf a})\approx A_{i}({\bf p})+
e a_{j}\partial A_{i}({\bf p})/\partial p_{j} $.
Taking this result into Eq. (9), then using the relation (11)
and integrating in parts, we rearrange $\gamma_{+}({\bf p})$ into:
\begin{equation}
\gamma_{+}({\bf p})=-\frac{e^{2}}{2}
\int\limits_{-\infty}^{+\infty} dt
~\left[\frac{p_{3}}{{\bf p}^{4}}
p_{k} \epsilon_{ijk}+\pi\delta(p_{1}, 
p_{2})\Delta(p_{3})\epsilon_{ij3}\right]a_{j} \dot{a}_{i}.
\end{equation}

Second step,  the geometric phase in the configuration
space is obtained through a Fourier transformation to 
$\gamma_{+}({\bf p})$:
\begin{equation}
\gamma_{+}({\bf x})=\frac{1}{(2\pi)^{3}}\int d^{3}{\bf p}
~\gamma_{+}({\bf p}) ~ e^{i{\bf p}\cdot {\bf x}}.
\end{equation}
Since there are infinite number of degrees of 
freedom in the system,  we should integrate over the 
whole space to get to the total contribution of the 
geometric phases to the gauge system:  
\begin{equation}
\Gamma_{+}=\int d^{3}{\bf x}~\gamma_{+}({\bf x}).
\end{equation}
$\Gamma_{+}$ is called the $geometric ~term$.

There are two remarks addressed in order:

(i) The above definition of $\Gamma_{+}$ 
is different from the definition of $\Gamma_{-}$ in Ref. [12],
where the electrons are restricted in two-dimensional Dirac sea,
thus have only the bounded states. 
However, the gauge system we consider is unbounded,
which can be realized by the low-energy scattering experiment
of electrons and photons. We notice that the Fourier transformation
was overlooked in Ref. [12], even though it does not yield
any difference between the result of Ref. [12] 
and that of ours as shown below. 

(ii) The second term in the expression of $\gamma_{+}({\bf p})$
is in fact a projection from three-dimensional space to 
two dimensions. It should be treated 
independently in $\Gamma_{+}$ through the two-dimensional 
Fourier
transformation:
\begin{equation}
\Gamma_{+}^{2}=\frac{1}{(2\pi)^{2}}
\int d t ~d^{2}{\bf x}\int d^{2}{\bf p}~
\left[\frac{-e^{2}}{2}\pi\delta(p_{1}, 
p_{2})\epsilon_{ij}a_{j} 
\dot{a}_{i}\right]~ e^{i{\bf p}\cdot {\bf x}},
\end{equation}
where $i,j=1,2$. $\Gamma_{+}^{2}$ is reduced to be:
\begin{equation}
\Gamma_{+}^{2}=-\frac{e^{2}}{8\pi}
\int  d t ~d^{2}{\bf x}~\epsilon_{ij}~a_{j} 
\dot{a}_{i},
\end{equation}
which is exactly the usual Chern-Simons action under the 
$temporal$ gauge condition: $a_{0}=0$.

We now deal with the remaining term in  $\Gamma_{+}$, which is
expressed as
\begin{equation}
\Gamma_{+}^{3}=\frac{ie^{2}}{16\pi^{3}}
\int d t ~d^{3}{\bf x}~\epsilon_{ijk}a_{j} \dot{a}_{i}
~\frac{\partial}{\partial x_{k}}f({\bf x}) ,
\end{equation} 
where
\begin{equation}
f({\bf x})=\int d^{3}{\bf p}~
\frac{p_{3}e^{i{\bf p}\cdot {\bf x}}}{{\bf p}^{4}}.
\end{equation}
It is evident that the above integral (18) is infrared divergent. 
Fortunately, this divergence can be resolved by
restoring the mass $m$ to the fermions, which leads to 
the replacement: ${\bf p}^{4}
\longmapsto~\left({\bf p}^{2}+
m^{2}\right)^{2}$ in the relevant term of Eq. (18). 
Then $f({\bf x}) \longmapsto f({\bf x}, m)$,
and  $\Gamma_{+}^{3}$ is  re-defined as: 
$\Gamma_{+}^{3}=\lim_{m\rightarrow 0}\Gamma_{+}^{3}(m)$.
A careful calculation gives neatly: $f({\bf x},m)=
i\pi^{2}~ e^{-m|{\bf x}|}$. Therefore,
\begin{equation}
\Gamma_{+}^{3}(m)=
\frac{-e^{2}}{16\pi}
\int d t ~d^{3}{\bf x}~\epsilon_{ijk}a_{j} \dot{a}_{i}
~\frac{\partial}{\partial x_{k}} e^{-m|{\bf x}|}.
\end{equation}
After integrating by parts  Eq. (19), we neglect the surface term
since: $e^{-m|{\bf x}|}
\rightarrow 0$, for $x_{k}\rightarrow \pm\infty$
($m$ is finite at this moment). Then choosing the limit
$m\rightarrow 0$, we get to:
\begin{equation}
\Gamma_{+}^{3}=
\frac{e^{2}}{16\pi}
\int d t ~d^{3}{\bf x}~\epsilon_{ijk} 
~\left[\partial_{k} a_{j} ~\dot{a}_{i}+ 
a_{j}~\partial_{k}
\dot{a}_{i}\right],
\end{equation}
which is further arranged into the following form 
after  integrating by parts the time derivative of the second term 
in $\Gamma_{+}^{3}$:
\begin{equation}
\Gamma_{+}^{3}=
-\frac{e^{2}}{8\pi}
\int d t ~d^{3}{\bf x}~\epsilon_{ijk} 
~\partial_{i} a_{j} ~\dot{a}_{k}.
\end{equation}
This $\Gamma_{+}^{3}$ is exactly
the well-known Pontrjagin term (precisely action) under the 
$temporal$ 
gauge condition: $a_{0}=0$, and the asymptotic behavior (5).
We notice that the usual  Pontrjagin term is a number-like term
with the factor $1/8\pi^{2}$. However, 
the above $\Gamma_{+}^{3}$  is a phase-like term,
thus has a different factor $\pi$ to the usual form. 
It should be pointed out that the above 
calculations can be extended to the non-Abelian 
gauge field interaction without difficulty.

The above results [Eqs. (16) and (21)] lead  to a conclusion
that  for the low-energy gauge interacting system,
the asymptotic behavior of the gauge field, based on the 
consideration of topological boundary, results in an additional
term to the effective action of the system: 
it is the Chern-Simons term for 
three-dimensional spacetime, and the Pontrjagin term for 
the four-dimensional spacetime. 
It is not an accident that the additional term---Chern-Simons term or 
Pontrjagin term, is a topological invariant,
because it comes from the Berry's argument of
geometric phase, which is associated with the nontrivial topological
structure---holonomy in the parameter space. 

According to the above argument of the role of geometric 
phase in the path integral formalism of quantum theory, we see that
in the functional formalism of gauge field theory, the geometric 
term $\Gamma_{+}^{2}$ or  $\Gamma_{+}^{3}$ can only appear
as an addition term in the $effective$ action, that is
\begin{equation}
S_{eff}=S +\Gamma_{+}^{i}, ~~~i=2,3,
\end{equation}
where $S$ is the usual ``classical" action as in Eq. (6).
One may wonder that since $\Gamma_{+}^{i}$ is a phase-type 
term, how it can enter the system as a part of 
$S_{eff}$ that determines the dynamical behavior of the system.
This could be answered from two considerations:
First, the term  $\Gamma_{+}^{i}$ essentially results from
the  Eq. (5), a topological requirement, even though it is
obtained by employing Berry's scheme. It is actually determined by 
the gauge field in functional form. Therefore, it can not 
be simply regarded as a phase. Instead, 
it takes into account the adiabatically evolution of the system from 
the point of view of low-energy perturbation. 
Second, we know that the action $S$ in Eq. (22) is the same 
a classical functional as $\Gamma_{+}^{i}$. Even if we treat
$S_{eff}$ as a quantum object to quantize it, as
many people have done, we see
that $\Gamma_{+}^{i}$ does not contribute to the 
Lagrangian equations of the quantum fields. 

We present several remarks on the above results:

(i)~ As is well known,  the topological term was first introduced 
in quantum gauge theory by resolving the chiral anomaly  in 
the one-loop triangle diagrams that breaks the usual
 Ward-Takahashi identities \cite{anomaly}.  Here we see that
the topological term, based on the low-energy 
approximation, enters the action automatically.
In field theory, it is not difficult to
extend the geometric phase from the classical quantity to 
the quantum quantity. Then the topological term, as
a quantum quantity, will be helpful in 
resolving the anomaly of the chiral current in gauge theory.
This will be discussed in the forthcoming paper.  

(ii) ~ The presence of Witten's work on topological field 
theory \cite{witten} has recently raised much more interest in 
theoretical physics. As that interpreted, there still is not 
a physical realization of the topological term as the dynamical 
term in action. Therefore, the functional integral  
of the topological field is called  partition function.
In our formulation, the topological term is interpreted 
as a result of the boundary condition of the gauge system,
in which the adiabatically condition is required.
We know that in the usual gauge field theory, 
the dynamics of the gauge system is determined  
by $S$ in Eq. (22). Now we may consider such 
an extreme situation that the system is confined in 
the ground state determined by $S$, 
which is highly degenerated and leads to vanish  of 
the average values of the dynamical terms. Then 
the topological term becomes to dominate the effective action.
In this consideration, the topological field theory could be 
regarded as the ground state reduction of the usual gauge 
field theory.  

(iii)~ It was revealed  recently that the Chern-Simons term
plays an important role in the low-dimensional physical world.
It brings two-dimensional nontrivial topology  \cite{tang} 
to the physical system,
and gives rise to some interesting observations, for
examples, exotic statistics of quasi-particles \cite{wilczek1}, 
the soliton solution \cite{jackiw}, and mass-generating of
the gauge field \cite{mass}.
Our formulation of action Eq. (22) implies that the above 
observations principally exist in the  gauge systems in
low dimensions, and the induced mass of the gauge field is 
determined by the coupling constant. The details are omitted here.  

In sum, we re-analyze the path integral formalism of the
gauge  system in this Letter.  
For the sufficiently low-energy gauge interaction,
the asymptotic behavior of the gauge field, 
based on the consideration of topological boundary, 
implies a closed loop traced by the gauge field  
in the time interval: $(-\infty,~+\infty)$.
Adopting Berry's argument of geometric phase, 
we show that the adiabatic evolution
of the gauge system around the loop results in an additional
term to the effective action: the Chern-Simons term for 
three-dimensional spacetime, and the Pontrjagin term for 
the four-dimensional spacetime. This approach
gives an alternative interpretation of the topological
terms in physics. \\

\acknowledgments

We thank F. T. Smith, S. Yu, C.L. Li
and I. Kulikov for their helpful discussions. 
The material is based upon research  supported in part by the
M\&H  Ferst Foundation, and by the NSF,  Grant No. PHY9211036.\\


\newpage

\begin{thebibliography}{s2}

\bibitem{anomaly}
J. Schwinger,  Phys. Rev. {\bf 82}, 664 (1951); 
S.L. Adler and W.R. Bardeen, 
Phys. Rev. {\bf 182}, 1517 (1969); J.S. Bell and 
R. Jackiw, Nuovo Cimento, {\bf
 60A}. 47 (1969).
 
\bibitem{instanton}
A.A. Belavin, A.M. Polyakov, 
A.S. Schwartz and Yu.S. Tyupkin, Phys. Lett.  {\bf 59B }, 85 (1975);
G. 't Hooft, Phys. Rev. D {\bf 14}, 3432 (1976).
 
\bibitem{witten}
E. Witten, Commun. Math. Phys. {\bf 117}, 353 (1988);
{\bf 121}, 351 (1989).

\bibitem{feynman}
P.R. Feynman and A.R. Hibbs (1964), $Quantum~mechanics 
~and ~path ~integrals$,
New York: McGraw-Hill.

\bibitem{dirac}
P.A.M. Dirac (1954), $Quantum~mechanics$, 4th
 ed.London: Oxford Univ. Press.

\bibitem{berry}
 M.V. Berry, Proc. Roy. Soc. London, Ser.A {\bf 392}, 45 (1984).

\bibitem{simon}
 B. Simon, Phys. Rev. Lett. {\bf 51}, 2167 (1983). 

\bibitem{nash}
C. Nash and S. Sen (1983), $Topology ~and ~ geometry~for ~
physicists$, London: Academic Press.

\bibitem{newton}
 R. G. Newton, Phys. Rev. Lett. {\bf 72}, 954 (1994).

\bibitem{wilczek}
 F. Wilczek and A. Zee, Phys. Rev. Lett. {\bf 52}, 2111 (1984).
 
\bibitem{nonabel}
The non-Abelian magnetic monopole is realized by 
$|\psi_{+}({\bf a})\rangle$ and $|\psi_{-}({\bf a})\rangle$ as 
two components, see S.N. Biswas, Phys. Lett. {\bf 228B}, 440 (1989).

\bibitem{li}
N. Fumita and K. Shizuya,  Phys. Rev. D{\bf 49}, 4277 (1994);
Z.S. Ma, S.S. Wu and H.Z. Li, Phys. Rev. D{\bf 52}, 337 (1995).

\bibitem{tang}
Z. Tang and D.R. Finkelstein,  $Connection~ between~ topology~ and
~statistics$, Reprint (1996).

\bibitem{wilczek1}
F. Wilczek (1990), 
$Fractional~ Statistics ~and ~anyon~ superconductivity$,
World Scitific  Pub. Co.

\bibitem{jackiw}
R. Jackiw and S.Y. Pi, 
 Phys. Rev. Lett. {\bf 64}, 2969 (1990).

\bibitem{mass}
G.E. Volovik, Sov. Phys. JETP, {\bf 67}, 1804 (1988);
V.M. Yakovenko,  Phys. Rev. Lett. {\bf 65}, 251 (1990).

\end{thebibliography}
\end{document}